\documentclass[uplatex]{interact}

\usepackage{epstopdf}
\usepackage{subfigure}
\usepackage{subfigmat}
\usepackage{hyperref,bm,amsmath,booktabs,multirow,color,here,graphicx}
\usepackage{caption}
\usepackage{natbib}
\bibpunct[, ]{(}{)}{;}{a}{}{,}

\newcommand{\argmin}{\mathop{\rm arg~min}\limits}

\theoremstyle{plain}

\theoremstyle{definition}

\theoremstyle{remark}

\begin{document}


\title{Bayesian Fused Lasso Modeling for Binary Data}

\author{
\name{Yuko Kakikawa\textsuperscript{a} and Shuichi Kawano\textsuperscript{b}\thanks{CONTACT Yuko Kakikawa. Email: kakikawa.yuko@ism.ac.jp}}
\affil{\textsuperscript{a} Department of Statistical Science, Graduate University of Advanced Studies (SOKENDAI), 10-3 Midori-cho, Tachikawa, Tokyo 190-8562, Japan.}\affil{\textsuperscript{b} Faculty of Mathematics, Kyushu University, 744 Motooka, Nishi-ku, Fukuoka 819-0395, Japan.}
}

\maketitle

\begin{abstract}
    $L_1$-norm regularized logistic regression models are widely used for analyzing data with binary response.
    In those analyses, fusing regression coefficients is useful for detecting groups of variables. 
    This paper proposes a binomial logistic regression model with Bayesian fused lasso. 
    Assuming a Laplace prior on regression coefficients and differences between adjacent regression coefficients enables us to perform variable selection and variable fusion simultaneously in the Bayesian framework. 
    We also propose assuming a horseshoe prior on the differences to improve the flexibility of variable fusion.
    The Gibbs sampler is derived to estimate the parameters by a hierarchical expression of priors and a data-augmentation method. 
    Using simulation studies and real data analysis, we compare the proposed methods with the existing method. 

\end{abstract}


\section{Introduction}
\label{sec:Introduction}
 Recently, $L_1$-norm regularized logistic regression models for binary response data have received considerable attention in various fields of research \citep{wu2009genome,pereira2016logistic,xin2014efficient,yu2015classification}. 
 The lasso \citep{tibshirani1996regression}, which is the most well-known $L_1$-norm regularized method, imposes the sum of absolute values of regression coefficients as a regularization term, and then estimates some regression coefficients as zeros. 
 This induces variable selection. 
 \citet{tibshirani1996regression} focused on the linear regression models, and subsequently many researchers have extended the lasso to logistic regression models (e.g., see \citet{lokhorst1999lasso}; \citet{shevade2003simple}; \citet{pereira2016logistic}).
Meanwhile, \citet{tibshirani2005sparsity} proposed the fused lasso, which not only performs variable selection but also estimates the differences between adjacent regression coefficients as exactly zero, that is, variable fusion. 
Through variable fusion, the group of variables which have similar roles in the prediction can be identified. 
\citet{yu2015classification} introduced a logistic regression model estimated by the fused lasso and applied it to spectral data.

From the perspective of Bayesian methods, a regularization term in the lasso
corresponds to assuming a Laplace distribution on the coefficients.
\citet{park2008bayesian} proposed the Bayesian lasso and enabled Gibbs sampling with the hierarchical expression of a Laplace prior by a scale mixture of normals \citep{andrews1974scale}. 
\citet{kyung2010penalized} extended the Bayesian lasso to the Bayesian fused lasso, which assumes a Laplace prior not only on regression coefficients but also on the differences of adjacent regression coefficients. 
\citet{genkin2007large} proposed the Bayesian lasso for logistic regression models for analyzing text data. 
\citet{betancourt2017bayesian} extended the Bayesian fused lasso to a logistic regression model, and then applied it to analyzing financial trading network data. 
However, these existing methods do not consider variable fusion and variable selection simultaneously. 

It should be noted that a Laplace prior induces over-shrinkage on the coefficients when the true coefficients are non-zero, while the prior induces under-shrinkage on the coefficients when the true coefficients are zero. 
This is a drawback to using a Laplace prior. 
As a counterpart of a Laplace prior, a horseshoe prior \citep{carvalho2010horseshoe} has been considered. 
A horseshoe prior has both a taller spike at zero and a heavier tail than a Laplace prior. 
These properties of this distribution overcome the above drawback of a Laplace prior. 
\citet{makalic2015simple} and \citet{bhattacharyya2022applications} proposed Bayesian logistic regression models by assuming a horseshoe prior on regression coefficients. 
However, the existing Bayesian logistic regression models with a horseshoe prior ignore variable fusion.

In the present paper, we first propose a Bayesian logistic regression model by assuming a Laplace prior on both regression coefficients and 
 differences between adjacent regression coefficients. 
 To prevent under-shrinkage and over-shrinkage of the differences, we next propose assuming a horseshoe prior on the differences between adjacent regression coefficients.
Subsequently, we derive the Gibbs sampler by using the hierarchical expression of a half-Cauchy distribution \citep{wand2011mean} and a data-augmentation method with a P\'{o}lya-Gamma distribution \citep{polson2013bayesian}.

The remainder of this paper is organized as follows. 
Section \ref{sec:L_1norm} introduces $L_{1}$-norm regularized logistic regression models. 
Section \ref{sec:Bayesian-regularized} describes a Bayesian approach for the $L_{1}$-norm regularized methods. 
In Section \ref{sec:proposed}, we present a Bayesian logistic regression model with both variable selection by a Laplace prior and variable fusion by a Laplace prior or a horseshoe prior.
Section \ref{sec:experiment} conducts Monte Carlo simulations and makes an application to real data to investigate the performance of our proposed methods, and compares it to the existing method, the logistic regression model with the fused lasso. 
We conclude the paper in Section \ref{sec:conclusion}.

\section{$L_{1}$-norm regularized logistic regression}
\label{sec:L_1norm}
Suppose that we have a dataset $\{ (y_i, \bm x_i); i=1,\ldots, n \}$, where $y_i \in \{0,1\}$ is a binary response variable and $\bm x_i$ is a $p$-dimensional vector of explanatory variables.
A logistic regression model is formulated as
\begin{equation}
	{\rm Pr}(y_{i}=1\mid\bm{x}_{i})
	=\frac{1}{1+e^{-(\beta_{0}+\bm{x}_{i}^{T}\bm{\beta})}},\quad i=1,\ldots,n,
	\label{eq:logistic}
	\end{equation}
where $\beta_{0}$ is an intercept and $\bm{\beta}=(\beta_{1},\beta_{2},\ldots,\beta_{p})^{T}$ is a $p$-dimensional regression coefficient vector. 
Then, the likelihood function is given by
\begin{equation}
	\prod_{i=1}^n f(y_{i}\mid\bm{x}_{i};\beta_{0},\bm{\beta}),
	\label{eq:yudo}
\end{equation}
where 
\begin{equation}
	f(y_{i}\mid\bm{x}_{i};\beta_{0},\bm{\beta})=\frac{(e^{\beta_{0}+\bm{x}_{i}^{T}\bm{\beta}})^{y_{i}}}{1+e^{\beta_{0}+\bm{x}_{i}^{T}\bm{\beta}}}.
	\label{eq:yudof}
\end{equation}
Then, the log-likelihood function is given by
	\begin{equation}
	\ell_{LR}(\beta_{0},\bm{\beta})=\sum_{i=1}^{n}\left[y_{i}(\beta_{0}+\bm{x}_{i}^{T}\bm{\beta})-\log{\{1+\exp({\beta_{0}+\bm{x}_{i}^{T}\bm{\beta}})\}}\right].\label{eq:taisuyudo}
	\end{equation}

In a maximum likelihood estimation, the regression coefficient $\bm{\beta}$ is estimated by the maximization of the log-likelihood \eqref{eq:taisuyudo}. 
However, the over-fitting problem often occurs when the sample size $n$ is smaller than the number of variables $p$. 
$L_{1}$-norm regularization is a useful tool for solving this problem. 
The lasso \citep{tibshirani1996regression,lokhorst1999lasso} is one of the most widely used $L_{1}$-norm regularization methods. 
The lasso is formulated as the following optimization problem:
\begin{equation}
	\hat{\bm{\beta}}_{L}=\argmin_{\beta_0, \bm{\beta}}\left\{-\ell_{LR}(\beta_{0},\bm{\beta})+\gamma\sum_{j=1}^{p}|\beta_{j}|\right\},
   \label{eq:lasso}
\end{equation}
where $\gamma \ (>0)$ is a regularization parameter that adjusts the strength of the shrinkage of regression coefficients. 
In the problem in \eqref{eq:lasso}, the sum of absolute values of regression coefficients is imposed as the regularization term, and thus the lasso estimates regression coefficients as exactly zero. 
Consequently, the lasso performs variable selection.

However, the lasso cannot consider the relationships among explanatory variables. 
When adjacent variables are strongly correlated, like in the case of spectral data \citep{yu2015classification}, adjacent variables have similar roles for the response.
For such data, regression coefficients which correspond to variables having similar roles should be estimated as the same values. 
To address this issue, the fused lasso \citep{tibshirani2005sparsity,yu2015classification} has been introduced. 
The fused lasso is formulated as the following minimization problem:
\begin{equation}
\hat{\bm{\beta}}_{FL}=\argmin_{\beta_0, \bm{\beta}}\left\{-\ell_{LR}(\beta_{0},\bm{\beta})+\lambda_{1}\sum_{j=1}^{p}|\beta_{j}|+\lambda_{2}\sum_{j=2}^{p}|\beta_{j}-\beta_{j-1}|\right\},
\label{eq:fusedlasso}
\end{equation}
where $\lambda_{1} \ (>0)$ and $\lambda_{2} \ (>0)$ are regularization parameters. 
The regularization parameter $\lambda_{1}$ adjusts the amount of the shrinkage of variables, while the regularization parameter $\lambda_{2}$ adjusts the smoothness of the differences between adjacent variables. 
The fused lasso \eqref{eq:fusedlasso} can estimate some of the regression coefficients and differences of adjacent regression coefficients as exactly zero.
This means that the fused lasso can perform variable selection and variable fusion simultaneously.

\section{Bayesian approach for regularized logistic regression}
\label{sec:Bayesian-regularized}
This section reviews a Bayesian logistic regression model and a data-augmentation method using a P\'{o}lya-Gamma distribution. 
We describe a Bayesian approach for $L_{1}$-norm regularized methods. 
Specifically, we explain the Bayesian lasso and the Bayesian fused lasso. 
We also describe a horseshoe prior and a Bayesian regularized logistic regression with a horseshoe prior.

\subsection{Data-augmentation method with P\'{o}lya-Gamma distribution}
\label{sec:Polya-Gamma distribution}
In the Bayesian framework, a logistic regression model is formulated as
\begin{equation}
	  y_{i} \mid \bm{x}_{i},\bm{\beta},\beta_{0}\sim {\rm Binom}\left(1,\frac{1}{1+e^{-(\beta_{0}+\bm{x}_{i}^{T}\bm{\beta})}}\right),
	  \label{logi_bayes}
\end{equation}
where ${\rm Binom}(\cdot,\cdot)$ represents a binomial distribution.

To obtain samples from the posterior distribution for \eqref{logi_bayes} by Gibbs sampling, \citet{polson2013bayesian} proposed a data-augmentation method with a P\'{o}lya-Gamma distribution. 
The probability density function of a P\'{o}lya-Gamma distribution is expressed by
  \begin{equation}
	{\rm PG}(x \mid a,b)=\frac{1}{2 \pi^2} \sum_{k=1}^{\infty} \frac{g_k}{ \left(k-\frac{1}{2}\right)^2+
	\frac{b^2}{(4 \pi^2)}},
	\label{eq:polyapdf}
  \end{equation}
  where $a \ (>0)$ and $b$ are hyper-parameters and $g_{k}$ is an independent random variable with a Gamma distribution ${\rm Ga}(a,1)$. 
  In addition, the P\'{o}lya-Gamma distribution \eqref{eq:polyapdf} can be expressed hierarchically with ${\rm PG}(x\mid a,0)$ as follows:
  \begin{equation}
	{\rm PG}(x \mid a,b)\propto \exp{\left(-\frac{b^{2}x}{2}\right)}{\rm PG}(x \mid a,0).
	\label{eq:polyapdfkaki}
  \end{equation}

  By using the hierarchical expression \eqref{eq:polyapdfkaki}, the function \eqref{eq:yudof} can be expressed hierarchically with the latent variables $w_{i}$:
  \begin{align}
	f(y_{i} \mid \bm{x}_{i},\bm{\beta},\beta_{0})&= \frac{(e^{\beta_{0}+\bm{x}_{i}^{T}\bm{\beta}})^{y_{i}}}{1+e^{\beta_{0}+\bm{x}_{i}^{T}\bm{\beta}}}\nonumber
	\\&
	=\frac{1}{2} \exp\{{\kappa_{i} (\beta_{0}+\bm{x}_{i}^{T}\bm{\beta})}\} \int_{0}^{\infty} \exp{\left\{-\frac{{w_{i} (\beta_{0}+\bm{x}_{i}^{T}\bm{\beta})}^{2}}{2}\right\}} {\rm PG}(w_{i} \mid 1, 0) d w_{i},
	\label{eq:polyalikelihood}
  \end{align}
  where $\kappa_{i}=y_{i}-0.5$. 
  When the regression coefficient vector $\bm{\beta}$ has a Gaussian prior ${\rm N}_p(\bm{0}_{p}, \bm{B})$, the full conditional distributions can be obtained as follows:
\begin{equation}
	\begin{split}
  \bm{\beta}\mid\bm{y},\bm{X},\beta_{0},\lambda_{1},w_{1},\ldots,w_{n}&\sim{\rm N}_{p}(\bm{A}^{-1}\bm{X}^{T}\bm{W}(\bm{z}-\beta_{0}\bm{1}),\bm{A}^{-1}),
  \\
  w_{i}\mid\bm{x}_{i},\bm{\beta},\beta_{0}&\sim {\rm PG}(1,\beta_{0}+\bm{x}_{i}^{T}\bm{\beta}),
  \label{eq:conditional_logi}
   \end{split}
\end{equation}
where $\bm{X}=(\bm{x}_{1},\bm{x}_{2},\ldots,\bm{x}_{n})^{T}$, $\bm{A}=\bm{X}^{T}\bm{W}\bm{X}+\bm{B}^{-1}$, $\bm{W}={\rm diag}(w_{1},\ldots,w_{n})$, $\bm{z}=(\kappa_{1}/w_{1},\kappa_{2}/w_{2},\ldots,\kappa_{n}/w_{n})^{T}$, and $\bm{1}$ is an $n$-dimensional vector of which all components are one. 
From the full conditional distributions \eqref{eq:conditional_logi}, the Gibbs sampling can be performed.

\subsection{Bayesian approach for $L_1$-norm regularized methods}
The lasso can be interpreted as a MAP estimation when a Laplace prior is assumed on regression coefficients independently. 
\citet{park2008bayesian} proposed the Bayesian lasso by assuming a Laplace prior ${\rm Laplace}(0,1/\lambda)$, where the location parameter is 0 and the scale parameter is $1/\lambda$, on regression coefficients in terms of linear regression models. 
\citet{park2008bayesian} also developed the Gibbs sampler by using a scale mixture of normals \citep{andrews1974scale}. 
With a scale mixture of normals, a Laplace prior can be expressed by a normal distribution and an exponential distribution hierarchically: 
  \begin{equation}
	\begin{split}
	  \bm{\beta}\mid \tau_{1}^{2},\ldots,\tau_{p}^{2}&\sim {\rm N}_{p}({\bm 0}_{p},\bm{C}),
	  \\\tau_{j}^{2}&\sim {\rm EXP}\left(\frac{\lambda^{2}}{2}\right),
	  \label{eq:lasso_logiseiki}
	\end{split}
  \end{equation}
  where $\bm{C}=$ diag $\displaystyle\left(\tau_{1}^{2},\tau_{2}^{2},\ldots,\tau_{p}^{2}\right)$  
  and ${\rm EXP}(\cdot)$ represents an exponential distribution.

  There are several ways to extend the Bayesian lasso in the framework of logistic regression models. 
  \citet{makalic2016high} proposed a logistic regression model with the priors \eqref{eq:lasso_logiseiki} in the case of $\lambda^{2}=2$ and $\bm{C}=$ diag $\displaystyle\left(\psi^{2}\tau_{1}^{2},\psi^{2}\tau_{2}^{2},\ldots,\psi^{2}\tau_{p}^{2}\right)$, where $\psi$ is a hyper-parameter which determines the overall shrinkage of the regression coefficients. 
  \citet{tian2019bayesian} also proposed a logistic regression model with the Bayesian lasso. 
  In \citet{tian2019bayesian}, the priors \eqref{eq:lasso_logiseiki} were assumed and the hyper-parameter $\lambda$ was selected to make a predictive probability distribution be an approximately uniform distribution.

  Among the models, we note the logistic regression model with the Bayesian lasso having priors \eqref{eq:lasso_logiseiki} on regression coefficients.
  First, as in \citet{makalic2016high}, a uniform distribution $\rm U(\alpha,\alpha)$ with a hyper-parameter $\alpha$ is assumed on the intercept $\beta_{0}$. In addition, a Gamma distribution ${\rm Ga} (r_1,\delta_1)\ (r_1>0,\ \delta_1>0)$ is assumed on $\lambda^{2}$, following \citet{park2008bayesian}.
  By combining the data-augmentation method in Sec. \ref{sec:Polya-Gamma distribution} and the expression for the prior on regression coefficients as a Gaussian scale-mixture prior \eqref{eq:lasso_logiseiki}, the full conditional distributions can be obtained as follows:
  \begin{equation}
  \begin{split}
  \bm{\beta}\mid\bm{y},\bm{X},\beta_{0},\tau_{1}^{2},\ldots,\tau_{p}^{2},w_{1},\ldots,w_{n}&\sim{\rm N}_{p}(\bm{A}^{-1}\bm{X}^{T}\bm{W}(\bm{z}-\beta_{0}\bm{1}),\bm{A}^{-1}),\nonumber
  \\
  w_{i}\mid\bm{x}_{i},\bm{\beta},\beta_{0}&\sim {\rm PG}(1,\beta_{0}+\bm{x}_{i}^{T}\bm{\beta}),\nonumber
  \\
  \frac{1}{\tau_{j}^{2}}\mid\beta_{j},\lambda^{2}&\sim{\rm IGauss}\left(\sqrt{\frac{\lambda^{2}}{\beta_{j}^{2}}},\lambda^{2}\right),\nonumber
  \\
  \lambda^{2}\mid\tau^{2}_{1},\ldots,\tau^{2}_{p}&\sim {\rm Ga}\left(p+r_{1},\frac{1}{2}\sum_{j=1}^{p}\tau_{j}^{2}+\delta_{1}\right),\nonumber
  \\
  \beta_{0}\mid \bm{y},\bm{X},\tau_{1}^{2},\ldots,\tau_{p}^{2},w_{1},\ldots,w_{n}&\sim {\rm N}\left(\frac{1}{S}\sum_{i=1}^{n}v_{i},\frac{1}{S}\right),\nonumber
  \end{split}
  \label{eq:conditional_lasso}
\end{equation}
where $\bm{A}=\bm{X}^{T}\bm{W}\bm{X}+\bm{C}^{-1}$, $\bm{v}=(\bm{z}-\bm{X\beta})\odot\bm{W}$, $S=\sum_{i=1}^{n}w_{i}$, 
and IGauss $(\cdot,\cdot)$ represents an inverse-Gaussian distribution and the symbol $\odot$ means the Hadamard product.

As an extension of the Bayesian lasso, \citet{kyung2010penalized} proposed a linear regression model which assumes a Laplace prior on regression coefficients and differences of adjacent regression coefficients. This extension is called the Bayesian fused lasso. The posterior mode of regression coefficients in the Bayesian fused lasso is equivalent to the fused lasso solution. 
\citet{betancourt2017bayesian} extended the Bayesian fused lasso to multinomial logistic regression by assuming a Laplace prior on the differences between parameters corresponding to adjacent points in a time series expressed by a directed binary network.

\subsection{Bayesian logistic regression model with horseshoe prior}

\label{sec:horseshoe}
A Laplace prior tends to shrink regression coefficients too little when the true coefficients are zero and shrink the coefficients too much when the true coefficients are non-zero. 
This is why a Laplace prior has insufficient concentration at zero and an exponential tail. 
To avoid this problem, \citet{carvalho2010horseshoe} proposed a horseshoe prior given by
\begin{equation}
	\beta_{j}|\lambda_{j}^{2},\tau^{2}\sim{\rm N}(0,\lambda_{j}^{2}\tau^{2}),\ \lambda_{j}\sim{\rm C^{+}}(0,1),\ \tau\sim{\rm C^{+}}(0,1),
	\label{eq:horseshoe}
\end{equation}
where ${\rm C}^{+}(\cdot,\cdot)$ represents a half-Cauchy distribution, $\lambda_{j}$ adjusts the degree of the shrinkage of each regression coefficient $\beta_{j}$, and $\tau$ adjusts the amount of the shrinkage of the overall regression coefficients $\bm{\beta}$. 
By having a hyper-parameter $\lambda_{j}$ following a half-Cauchy distribution with a pole at zero and a polynomial tail, a horseshoe prior realizes an infinite spike at zero and a heavier tail than a Laplace prior. Therefore, a horseshoe prior can strongly shrink the small regression coefficients towards zero and prevent over-shrinkage of large regression coefficients.

\citet{makalic2015simple} proposed a logistic regression model whose regression coefficients follow a horseshoe prior.
They utilized the hierarchical expression of a half-Cauchy distribution \citep{wand2011mean}. 
When $x$ follows ${\rm C}^{+}(0,1)$, the relation
\begin{equation}
x^{2} \mid y\sim{\rm IG}\left(\frac{1}{2},\frac{1}{y}\right),\ \ \ y\sim{\rm IG}\left(\frac{1}{2},1\right)\nonumber
\label{eq:halfcauchyxa}
\end{equation}
holds, 
where ${\rm IG}(\cdot,\cdot)$ represents an inverse-Gamma prior. 
 Therefore, the priors \eqref{eq:horseshoe} can be rewritten as follows:
 \begin{equation}
	\begin{split}
	  \bm{\beta}\mid \lambda_{1}^{2},\ldots,\lambda_{p}^{2},\tau^{2}&\sim {\rm N}_{p}({\bm 0}_{p},\bm{D}),
	  \\
	\lambda_{j}^{2} \mid \nu_{j}&\sim{\rm IG}\left(\frac{1}{2},\frac{1}{\nu_{j}}\right),
	\\
	\tau^{2}\mid\xi&\sim{\rm IG}\left(\frac{1}{2},\frac{1}{\xi}\right),
	\\
	\nu_{1},\ldots,\nu_{p},\xi&\sim\rm{IG}\left(\frac{1}{2},1\right),
	\end{split}
	\label{eq:makalicpriorkaki}
	\end{equation}
	where $\bm{D}=$ diag
	 $\displaystyle\left(\tau^{2}{{\lambda_{1}^{2}}},\tau^{2}{{\lambda_{2}^{2}}}\ldots,\tau^{2}{\lambda_{p}^{2}}\right)$.
 By using the priors \eqref{eq:makalicpriorkaki} and the data-augmentation method with a P\'{o}lya-Gamma distribution, the full conditional distributions can be obtained. The details are found in \citet{makalic2015simple}.

\section{Proposed methods}
\label{sec:proposed}
In this section, we propose a Bayesian logistic regression model whose regression coefficients and differences between adjacent coefficients follow a Laplace prior. 
We also propose an alternative Bayesian logistic regression model with a horseshoe prior on the differences.

\subsection{Logistic regression model with the Bayesian fused lasso}
\label{sec:ProposedBayesianFusedLasso}
We propose a logistic regression model which assumes a Laplace prior on regression coefficients and differences between adjacent regression coefficients.

First, we assume the following priors:
\begin{equation}
\begin{split}
\beta_{j} \mid \lambda_{1}&\sim {\rm Laplace}\left(0,\frac{1}{\lambda_{1}}\right),
\\
\beta_{j}-\beta_{j-1} \mid \lambda_{2}&\sim {\rm Laplace}\left(0,\frac{1}{\lambda_{2}}\right).
\end{split}
\label{eq:bayesianfusedlasso}
\end{equation}
The priors \eqref{eq:bayesianfusedlasso} induce shrinkage of both regression coefficients and differences of adjacent regression coefficients towards zero.
By using a scale mixture of normals, the priors \eqref{eq:bayesianfusedlasso} can be expressed as follows:
\begin{equation}
	\begin{split}
	  \\\bm{\beta}\mid \tau_{1}^{2},\ldots,\tau_{p}^{2},\tilde{\tau}_{2}^{2},\ldots,\tilde{\tau}_{p}^{2}&\sim {\rm N}_{p}({\bm 0}_{p},\bm{E}),
	  \\\tau_{j}^{2}&\sim {\rm EXP}\left(\frac{\lambda_{1}^{2}}{2}\right),
	  \\\tilde{\tau}_{j}^{2}&\sim {\rm EXP}\left(\frac{\lambda_{2}^{2}}{2}\right),
	  \label{eq:bayesfusedlassoseiki}
	\end{split}
  \end{equation}
  where the inverse matrix of $\bm{E}$ is expressed as
  \begin{equation}
  \bm{E}^{-1}=\left(\begin{array}{cccccc}
  \frac{1}{\tau_{1}^{2}}+\frac{1}{\tilde{\tau}_{2}^{2}}&-\frac{1}{\tilde{\tau}_{2}^{2}}&0&\ldots&0&0\\
  -\frac{1}{\tilde{\tau}_{2}^{2}}&\frac{1}{\tau_{2}^{2}}+\frac{1}{\tilde{\tau}_{2}^{2}}+\frac{1}{\tilde{\tau}_{3}^{2}}&-\frac{1}{\tilde{\tau}_{3}^{2}}&\ldots&0&0\\
  0&-\frac{1}{\tilde{\tau}_{3}^{2}}&\frac{1}{\tau_{3}^{2}}+\frac{1}{\tilde{\tau}_{3}^{2}}+\frac{1}{\tilde{\tau}_{4}^{2}}&\ldots&0&0\\
  \vdots&\vdots&\vdots&\ddots&\vdots&\vdots\\
  0&0&0&\ldots&\frac{1}{\tau_{p-1}^{2}}+\frac{1}{\tilde{\tau}_{p-1}^{2}}+\frac{1}{\tilde{\tau}_{p}^{2}}&-\frac{1}{\tilde{\tau}_{p}^{2}}\\
  0&0&0&\ldots&-\frac{1}{\tilde{\tau}_{p}^{2}}&\frac{1}{\tau_{p}^{2}}+\frac{1}{\tilde{\tau}_{p}^{2}}
  \end{array}\right).\nonumber
  \end{equation}
We assume priors on the intercept and hyper-parameters as follows:
\begin{equation}
\begin{split}
\beta_{0}&\sim {\rm U}(\alpha,\alpha),\\
\lambda_{1}^2&\sim {\rm Ga}(r_{1},\delta_{1}),\\
\lambda_{2}^2&\sim {\rm Ga}(r_{2},\delta_{2}),
\label{eq:hyperparamter}
\end{split}
\end{equation}
where $r_{1}, r_{2},\delta_{1}$, and $\delta_{2}$ are positive parameters. 
The priors \eqref{eq:hyperparamter} enable fully Bayesian estimation. 
This formulation can be regarded as an extension of the Bayesian fused lasso to a logistic regression model.

By combining a data-augmentation method with a P\'{o}lya-Gamma distribution and the hierarchical expression of the priors \eqref{eq:bayesfusedlassoseiki}, the full conditional distributions can be obtained as follows:
\begin{align}
	\bm{\beta}\mid\bm{y},\bm{X},\beta_{0},\tau_{1}^{2},\ldots,\tau_{p}^{2},\tilde{\tau}_{2}^{2},\ldots,\tilde{\tau}_{p}^{2},w_{1},\ldots,w_{n}&\sim{\rm N}_{p}(\bm{A}^{-1}\bm{X}^{T}\bm{W}(\bm{z}-\beta_{0}\bm{1}),\bm{A}^{-1}),\nonumber
	\\
	w_{i}\mid\bm{x}_{i},\bm{\beta},\beta_{0}&\sim {\rm PG}(1,\beta_{0}+\bm{x}_{i}^{T}\bm{\beta}),\nonumber
	\\
	\frac{1}{\tau_{j}^{2}}\mid\beta_{j},\lambda_{1}^{2}&\sim{\rm IGauss}\left(\sqrt{\frac{\lambda_{1}^{2}}{\beta_{j}^{2}}},\lambda_{1}^{2}\right),\nonumber
	\\
	\lambda_{1}^{2}\mid\tau^{2}_{1},\ldots,\tau^{2}_{p}&\sim {\rm Ga}\left(p+r_{1},\frac{1}{2}\sum_{j=1}^{p}\tau_{j}^{2}+\delta_{1}\right),\nonumber
	\\
	\frac{1}{\tilde{\tau}_{j}^{2}}\mid\beta_{j},\beta_{j-1}, \lambda_{2}^{2}&\sim{\rm IGauss}\left(\sqrt{\frac{\lambda_{2}^{2}}{(\beta_{j}-\beta_{j-1})^{2}}},\lambda_{2}^{2}\right),\nonumber
	\\
	\lambda_{2}^{2}\mid\tilde{\tau}^{2}_{2},\ldots,\tilde{\tau}^{2}_{p}&\sim {\rm Ga}\left(p-1+r_{2},\frac{1}{2}\sum_{j=2}^{p}\tilde{\tau}_{j}^{2}+\delta_{2}\right),\nonumber
	\\
	\beta_{0}\mid \bm{y},\bm{X},\bm{\beta},\tau_{1}^{2},\ldots,\tau_{p}^{2},\tilde{\tau}_{2}^{2},\ldots,\tilde{\tau}_{p}^{2},w_{1},\ldots,w_{n}&\sim {\rm N}\left(\frac{1}{S}\sum_{i=1}^{n}v_{i},\frac{1}{S}\right),
	\label{eq:conditional_fused}
  \end{align}
  where $\bm{A}=\bm{X}^{T}\bm{W}\bm{X}+\bm{E}^{-1}$, $\bm{v}=(\bm{z}-\bm{X\beta})\odot\bm{W}$, and $S=\sum_{i=1}^{n}w_{i}$. 
  From these full conditional distributions, Gibbs sampling can be performed.

\subsection{Logistic regression model with Bayesian fused lasso with horseshoe prior}
We also propose an alternative Bayesian model in Sec. \ref{sec:ProposedBayesianFusedLasso} by assuming a horseshoe prior on differences between adjacent regression coefficients.

First, we introduce the priors given by
\begin{equation}
   \begin{split}
\beta_{j} \mid \tilde{\lambda}_{1}&\sim {\rm Laplace}\left(0,\frac{1}{\tilde{\lambda}_{1}}\right),
   \\\beta_{j}-\beta_{j-1} \mid \lambda_{j}^{2},\tilde{\tau}^{2}&\sim
   {\rm N}\left(0,\lambda_{\textit{j}}^{2}\tilde{\tau}^{2}\right),
   \\
   \lambda_{j}&\sim {\rm C}^{+}(0,1),
   \\
   \tilde{\tau}&\sim  {\rm C}^{+}(0,1).
   \end{split}
   \label{eq:proposehorse}
\end{equation}
With the hierarchical expressions of a Laplace distribution and a half-Cauchy distribution, the priors \eqref{eq:proposehorse} are expressed as
\begin{equation}
	\begin{split}
		\bm{\beta}\mid \tau_{1}^{2},\ldots,\tau_{p}^{2},\lambda_{2}^{2},\ldots,\lambda_{p}^{2},\tilde{\tau}^{2}&\sim {\rm N}_{p}({\bm 0}_{p},\bm{F}),
		\\
		\tau_{j}^{2}&\sim{\rm EXP}\left(\frac{\tilde{\lambda}_{1}^{2}}{2}\right),
		\\
		\tilde{\tau}^{2}\mid\xi&\sim{\rm IG}\left(\frac{1}{2},\frac{1}{\xi}\right),
		\\
		\lambda_{j}^{2}\mid\nu_{j}&\sim{\rm IG}\left(\frac{1}{2},\frac{1}{\nu_{j}}\right),
		\\
		\xi,\nu_{j}&\sim{\rm IG}\left(\frac{1}{2},1\right),
		\end{split}
		\label{eq:yukomodelkaki}
		\end{equation}
		where the inverse matrix of $\bm{F}$ is given by
		\begin{equation}
		  \bm{F}^{-1}=\left(\begin{array}{cccccc}
		  \frac{1}{\tau_{1}^{2}}+\frac{1}{\lambda_{2}^{2}\tilde{\tau}^{2}}&-\frac{1}{\lambda_{2}^{2}\tilde{\tau}^{2}}&0&\ldots&0&0\\
		  -\frac{1}{\lambda_{2}^{2}\tilde{\tau}^{2}}&\frac{1}{\tau_{2}^{2}}+\frac{1}{\lambda_{2}^{2}\tilde{\tau}^{2}}+\frac{1}{\lambda_{3}^{2}\tilde{\tau}^{2}}&-\frac{1}{\lambda_{3}^{2}\tilde{\tau}^{2}}&\ldots&0&0\\
		  0&-\frac{1}{\lambda_{3}^{2}\tilde{\tau}^{2}}&\frac{1}{\tau_{3}^{2}}+\frac{1}{\lambda_{3}^{2}\tilde{\tau}^{2}}+\frac{1}{\lambda_{4}^{2}\tilde{\tau}^{2}}&\ldots&0&0\\
		  \vdots&\vdots&\vdots&\ddots&\vdots&\vdots\\
		  0&0&0&\ldots&\frac{1}{\tau_{p-1}^{2}}+\frac{1}{\lambda_{p-1}^{2}\tilde{\tau}^{2}}+\frac{1}{\lambda_{p}^{2}\tilde{\tau}^{2}}&-\frac{1}{\lambda_{p}^{2}\tilde{\tau}^{2}}\\
		  0&0&0&\ldots&-\frac{1}{\lambda_{p}^{2}\tilde{\tau}^{2}}&\frac{1}{\tau_{p}^{2}}+\frac{1}{\lambda_{p}^{2}\tilde{\tau}^{2}}
		  \end{array}\right).\nonumber
		\end{equation}
	By assuming a horseshoe prior on differences between adjacent regression coefficients, small differences can be shrunk more, while large differences are shrunk less, compared to a Laplace prior.

		From the priors \eqref{eq:yukomodelkaki} and the data-augmentation method, we can get the full conditional distributions as follows:
		\begin{equation}
		  \begin{split}
			\bm{\beta}\mid\bm{y},\bm{X},\beta_{0},\tau_{1}^{2},\ldots,\tau_{p}^{2},\lambda_{2}^{2},\ldots,\lambda_{p}^{2},\tilde{\tau}^{2},w_{1},\ldots,w_{n}&\sim{\rm N}_{p}(\bm{A}^{-1}\bm{X}^{T}\bm{W}(\bm{z}-\beta_{0}\bm{1}),\bm{A}^{-1}),
			\\
			w_{i}\mid\bm{x}_{i},\bm{\beta},\beta_{0}&\sim {\rm PG}(1,\beta_{0}+\bm{x}_{i}^{T}\bm{\beta}),
			\\
			\frac{1}{\tau_{j}^{2}}\mid\beta_{j},\tilde{\lambda}_{1}^{2}&\sim{\rm IGauss}\left(\sqrt{\frac{\tilde{\lambda}_{1}^{2}}{\beta_{j}^{2}}},\tilde{\lambda}_{1}^{2}\right),
			\\
			\tilde{\lambda}_{1}^{2}\mid\tau^{2}_{1},\ldots,\tau^{2}_{p}&\sim {\rm Ga}\left(p+r_{1},\frac{1}{2}\sum_{j=1}^{p}\tau_{j}^{2}+\delta_{1}\right),\nonumber
		  \end{split}
			\end{equation}
			\begin{equation}
			  \begin{split}
			\tilde{\tau}^{2}\mid\beta_{1},\ldots,\beta_{p},\lambda_{2}^{2},\ldots,\lambda_{p}^{2},\xi&\sim{\rm IG}\left(\frac{p}{2},\frac{1}{2}\sum^{p}_{j=2}\frac{(\beta_{j}-\beta_{j-1})^{2}}{\lambda_{j}^{2}}+\frac{1}{\xi}\right),\nonumber
			\\
			\lambda_{j}^{2}\mid\beta_{j},\beta_{j-1},\tilde{\tau}^{2},\nu_{j}&\sim{\rm IG}\left(1,\frac{(\beta_{j}-\beta_{j-1})^{2}}{2\tilde{\tau}^{2}}+\frac{1}{\nu_{j}}\right),
			\\
			\nu_{j}\mid\lambda_{j}^{2}&\sim{\rm IG}\left(1,\frac{1}{\lambda_{j}^{2}}+1\right),
			\\
			\xi\mid\tilde{\tau}^{2}&\sim{\rm IG}\left(1,\frac{1}{\tilde{\tau}^{2}}+1\right),
		  \\
			\beta_{0}\mid \bm{y},\bm{X},\tau_{1}^{2},\ldots,\tau_{p}^{2},\lambda_{2}^{2},\ldots,\lambda_{p}^{2},\tilde{\tau}^{2},w_{1},\ldots,w_{n}&\sim {\rm N}\left(\frac{1}{S}\sum_{i=1}^{n}v_{i},\frac{1}{S}\right),
			\label{eq:yukomodel_condi}
		  \end{split}
		\end{equation}		
where $\bm{A}=\bm{X}^{T}\bm{W}\bm{X}+\bm{F}^{-1}$, $\bm{v}=(\bm{z}-\bm{X\beta})\odot\bm{W}$, and $S=\sum_{i=1}^{n}w_{i}$.

Note that we did not assume a horseshoe prior on regression coefficients, because the MCMC chain did not converge.


\section{Numerical studies}
\label{sec:experiment}
In this section, we describe the settings and results of Monte Carlo simulations and compare the performance of our proposed methods with the existing method. 
We also apply our proposed methods to anomaly detection in the field of semiconductor microelectronics manufacturing \citep{olszewski2001generalized,deng2014smt}. 

\subsection{Monte Carlo simulations}
\label{sec:Monte Carlo}

We generated $y_i \ (i=1,2,\ldots,n)$ according to the true model:
\begin{equation}
	{\rm Pr}(y_{i}=1\mid\bm{x}_{i})=\displaystyle\frac{1}{1+e^{-\bm{x}_{i}^{T}\bm{\beta}^{*}}},
  \end{equation}
where $\bm{\beta}^{*}=(\beta_{1}^{*},\beta_{2}^{*},\ldots, \beta_{p}^{*})^{T}$ is a $p$-dimensional regression coefficient vector. 
The explanatory variable $\bm{x}_{i}\ (i=1,2,\ldots,n)$ followed the multivariate normal distribution ${\rm N}_{p}(\bm{0}_p, \Sigma)$.
For $\bm \beta^\ast$ and $\Sigma$, we considered the following cases:\\
\newline
Case 1: $\bm{\beta}^{*}=\bm{\beta}_{1}^{*}$ or $\bm{\beta}_{2}^{*}$, $\Sigma_{ii}=1$, $\Sigma_{ij}=\rho,\hspace{12pt}(i\neq j)$,\\
Case 2: $\bm{\beta}^{*}=\bm{\beta}_{1}^{*}$ or $\bm{\beta}_{2}^{*}$,\\
\vspace{0pt}
\hspace{28pt}$\Sigma_{ii}=1$, $\Sigma_{ij}=
\begin{cases}
 0.5&(\beta^{*}_i=\beta^{*}_{j}\mbox{ and } 1\leq |i-j|\leq 4)\\
 0&\mbox{otherwise}
\end{cases}$$\hspace{12pt}(i\neq j)$,\\
Case 3: $\bm{\beta}^{*}=\bm{\beta}_{1}^{*}$ or $\bm{\beta}_{2}^{*}$,\\
\vspace{0pt}
\hspace{28pt}$\Sigma_{ii}=1$, $\Sigma_{ij}=
\begin{cases}
 0.5^{|i-j|}&(\beta^{*}_{i}=\beta^{*}_{j}\mbox{ and } 1\leq |i-j|\leq 4)\\
 0&\mbox{otherwise}
\end{cases}$$\hspace{12pt}(i\neq j)$,\\
Case 4: $\bm{\beta}^{*}=(\bm{1.0}_{20}^{T}, \bm{-1.0}_{20}^{T}, \bm{0.0}_{170}^{T}, \bm{1.5}_{20}^{T},\bm{0.0}_{170}^{T})^{T}$, $\Sigma_{ii}=1$, $\Sigma_{ij}=0,\ \hspace{12pt}(i\neq j)$,\\
\newline
where $\bm{\beta}_{1}^{*}=(\bm{1.0}_{5}^{T}, \bm{0.0}_{5}^{T}, \bm{1.0}_{5}^{T},  \bm{0.0}_{5}^{T})^{T}$, $\bm{\beta}_{2}^{*}=(\bm{-1.0}_{5}^{T}, \bm{2.0}_{5}^{T}, \bm{1.0}_{5}^{T},  \bm{0.0}_{5}^{T})^{T}$, $\Sigma_{ij}$ is the $(i,j)$-th element of $\Sigma$, 
 and $\rho=0.0, 0.5$. 
We considered $n=500$ for Cases 1, 2, and 3, which correspond to $n>p$ cases, while $n=300$ for Case 4, which corresponds to an $n<p$ case.
We simulated 100 datasets for each case.
Note that Case 4 is based on the simulation in \citet{bhattacharyya2022applications}.

We compared our proposed methods, which are the logistic regression model with the Bayesian fused lasso (LBFL) and that with the Bayesian fused lasso with horseshoe prior (LBFH), to the logistic regression model with the fused lasso (LFL). 
For LFL, we used the package \textsf{penalized} of the software \textbf{R}, which is available from \url{https://cran.r-project.org/web/packages/penalized/index.html}. 
The values of the hyper-parameters $\lambda_1$ and $\lambda_2$ for LFL were selected by the Bayesian information criterion (BIC). 
As the term of the degrees of freedom in BIC, we used the number of groups which consist of non-zero fused estimated regression coefficients \citep{tibshirani2005sparsity,tibshirani2011solution}. 
For LBFL and LBFH, the Gibbs sampling was run with 10,000 iterations, and then the first 6,000 iterations were discarded as burn-in.

To measure the accuracy of the estimation of regression coefficients, we computed the mean squared error (MSE):
\begin{equation}
 {\rm MSE}=\frac{1}{100}\sum_{k=1}^{100} \left[ (\hat{\beta}_0^{(k)})^2 + \left\{{\hat{\bm{\beta}}}^{(k)}-\bm{\beta}^{*}\right\}^{T}\left\{{\hat{\bm{\beta}}}^{(k)}-\bm{\beta}^{*}\right\} \right],
\end{equation}
where $\hat{\beta}_{0}^{(k)}$ and ${\hat{\bm{\beta}}}^{(k)}=(\hat{\beta}_{1}^{(k)},\ldots,\hat{\beta}_{p}^{(k)})^{T}$ are an intercept and a vector of regression coefficients estimated from the $k$-th dataset, respectively. 
To evaluate the prediction accuracy, we used the negative expected log-likelihood. 
We generated 1,000 test data, and then computed the mean of the empirical negative expected log-likelihood:
\begin{equation}
  {\rm EL}=\frac{1}{100}\sum_{k=1}^{100}\left\{-\frac{1}{1000}\sum_{j=1}^{1000}\left[{y_{j}^{\dagger}}^{(k)}({{\bm{x}^{\dagger}}^{(k)}_{j}}^{T}\hat{\bm{\beta}}^{(k)}+\hat{\beta}_{0}^{(k)})-\log\{1+\exp(\hat{\bm{\beta}}^{(k)}+\hat{\beta}_{0}^{(k)})\}\right]\right\},
\end{equation}
where ${y_{j}^{\dagger}}^{(k)}$ and ${\bm{x}^{\dagger}}^{(k)}_{j}$ are the $j$-th data in the test data for the $k$-th dataset. 

To assess the performance of variable selection and variable fusion, we computed the following measures:
\begin{equation*}
\begin{split}
{\rm PV}&=\frac{1}{100}\sum_{k=1}^{100}\frac{|\{j \mid ( \beta^{*}_{j}\neq0 ) \land ( 0 \notin {\rm CIV}_{j}(k))\}|}{|\{j\mid\beta^{*}_{j}\neq0\}|}\hspace{12pt}(1\leq j\leq p),\\
{\rm PZV}&=\frac{1}{100}\sum_{k=1}^{100}\frac{|\{j\mid(\beta^{*}_{j}=0) \land (0\in {\rm CIV}_{j}(k))\}|}{|\{j\mid\beta^{*}_{j}=0\}|}\hspace{12pt}(1\leq j\leq p),\\
{\rm AV}&=\frac{1}{100}\sum_{k=1}^{100}\frac{|\{j\mid(\beta^{*}_{j}\neq0) \land ({0\notin\rm CIV}_{j}(k))\}|+|\{j\mid(\beta^{*}_{j}=0) \land (0 \in {\rm CIV}_{j}(k))\}|}{p}\hspace{12pt}(1\leq j\leq p),\\
{\rm PF}&=\frac{1}{100}\sum_{k=1}^{100}\frac{|\{j\mid(\beta^{*}_{j}-\beta^{*}_{j-1}\neq0) \land (0\notin{\rm CIF}_{j}(k))\}|}{|\{j\mid\beta^{*}_{j}-\beta^{*}_{j-1}\neq0\}|}\hspace{12pt}(2\leq j\leq p),\\
{\rm PNF}&=\frac{1}{100}\sum_{k=1}^{100}\frac{|\{j\mid(\beta^{*}_{j}-\beta^{*}_{j-1}=0) \land (0\in{\rm CIF}_{j}(k))\}|}{|\{j\mid\beta^{*}_{j}-\beta^{*}_{j-1}=0\}|}\hspace{12pt}(2\leq j\leq p),\\
{\rm AF}&=\frac{1}{100}\sum_{k=1}^{100}\frac{|\{j \mid (\beta^{*}_{j}-\beta^{*}_{j-1}\neq0) \land (0\notin{\rm CIF}_{j}(k))\}|+|\{j \mid (\beta^{*}_{j}-\beta^{*}_{j-1}=0) \land (0\in{\rm CIF}_{j}(k))\}|}{p-1}\\
&\hspace{350pt}(2\leq j\leq p),
\end{split}
\end{equation*}
where ${\rm CIV}_{j}(k)$ denotes the 95\% credible interval for $\hat{\beta}_{j}^{(k)}$,
 while ${\rm CIF}_{j}(k)$ denotes the 50\% credible interval for the difference between $\hat{\beta}_{j}^{(k)}$ and $\hat{\beta}_{j-1}^{(k)}$. 
PV and PZV measure the accuracy of estimating regression coefficients when the corresponding true regression coefficients are non-zero and zero, respectively. 
PF and PNF measure the accuracy of the differences between estimated adjacent regression coefficients when the differences between the true adjacent regression coefficients are non-zero and zero, respectively.
AV and AF measure the accuracy of variable selection and variable fusion, respectively.

Note that we utilized credible intervals of MCMC samples in evaluating the performance of variable selection and variable fusion. 
The reason why LBFL and LBFH do not estimate regression coefficients and their differences as exactly zero is that the posteriors of them are continuous. 
For variable selection, we computed 95\% credible intervals as in \citet{bhattacharyya2022applications}, and then judged that the regression coefficient is estimated as zero when the credible interval of its MCMC samples includes zero. 
For variable fusion, we computed 50\% credible intervals of the differences between adjacent regression coefficients as in \citet{banerjee2022horseshoe}, and then judged that the difference is estimated as zero, similar to variable selection.

\begin{table}[H]
   	\centering
      \caption{MSE (standard deviation), EL, PV, PZV, AV, PF, PNF, and AF for Case 1 and $\rho=0$. Bold font indicates the smallest value of MSE and EL and the largest value of PV, PZV, AV, PF, PNF, and AF before rounding among LFL, LBFL, and LBFH.}
      \label{tab:case1-0}
   	\begin{tabular}{c|ccccccccc}
   	\hline
      \multicolumn{2}{c}{}&MSE&EL&PV&PZV&AV&PF&PNF&AF\\
      \multicolumn{2}{c}{}&(sd)&(sd)&(sd)&(sd)&(sd)&(sd)&(sd)&(sd)\\
      \hline
      \multirow{6}{*}{$\bm{\beta}_{1}^{*}$}&\multirow{2}{*}{LFL}&0.619&180.551&\textbf{1.000}&0.501&0.751&\textbf{1.000}&0.766&0.803\\
      &&(0.287)&(3.032)&\textbf{(0.000)}&(0.288)&(0.144)&\textbf{(0.000)}&(0.113)&(0.095)\\
      &\multirow{2}{*}{LBFL}&0.490&183.965&\textbf{1.000}&0.911&0.956&\textbf{1.000}&0.541&0.613\\
      &&(0.303)&(4.997)&\textbf{(0.000)}&(0.094)&(0.047)&\textbf{(0.000)}&(0.143)&(0.120)\\
      &\multirow{2}{*}{LBFH}&\textbf{0.252}&\textbf{178.379}&\textbf{1.000}&\textbf{0.961}&\textbf{0.981}&\textbf{1.000}&\textbf{0.850}&\textbf{0.874}\\
      &&\textbf{(0.167)}&\textbf{(3.135)}&\textbf{(0.000)}&\textbf{(0.076)}&\textbf{(0.038)}&\textbf{(0.000)}&\textbf{(0.105)}&\textbf{(0.089)}\\
      \hline
      \multirow{6}{*}{$\bm{\beta}_{2}^{*}$}&\multirow{2}{*}{LFL}&3.722&123.790&\textbf{1.000}&0.480&0.870&\textbf{1.000}&0.769&0.805\\
      &&(0.898)&(3.398)&\textbf{(0.000)}&(0.417)&(0.104)&\textbf{(0.000)}&(0.101)&(0.085)\\
      &\multirow{2}{*}{LBFL}&0.905&122.723&\textbf{1.000}&0.944&0.986&\textbf{1.000}&0.541&0.613\\
      &&(0.494)&(3.846)&\textbf{(0.000)}&(0.107)&(0.027)&\textbf{(0.000)}&(0.135)&(0.114)\\
      &\multirow{2}{*}{LBFH}&\textbf{0.809}&\textbf{117.895}&\textbf{1.000}&\textbf{0.972}&\textbf{0.993}&\textbf{1.000}&\textbf{0.868}&\textbf{0.888}\\
      &&\textbf{(0.522)}&\textbf{(3.084)}&\textbf{(0.000)}&\textbf{(0.103)}&\textbf{(0.026)}&\textbf{(0.000)}&\textbf{(0.083)}&\textbf{(0.069)}\\
      \hline
   	\end{tabular}
      \end{table}
   
   \begin{table}[H]
   	\centering
   	\caption{MSE (standard deviation), EL, PV, PZV, AV, PF, PNF, and AF for Case 1 and $\rho=0.5$. Bold font indicates the smallest value of MSE and EL and the largest value of PV, PZV, AV, PF, PNF, and AF before rounding among LFL, LBFL, and LBFH.}
      \label{tab:case1-0.5}
      \begin{tabular}{c|ccccccccc}
   	\hline
      \multicolumn{2}{c}{}&MSE&EL&PV&PZV&AV&PF&PNF&AF\\
      \multicolumn{2}{c}{}&(sd)&(sd)&(sd)&(sd)&(sd)&(sd)&(sd)&(sd)\\
      \hline
      \multirow{6}{*}{$\bm{\beta}_{1}^{*}$}&\multirow{2}{*}{LFL}&1.735&94.449&\textbf{1.000}&0.236&0.618&0.940&\textbf{0.901}&\textbf{0.907}\\
      &&(0.634)&(4.134)&\textbf{(0.000)}&(0.261)&(0.130)&(0.137)&\textbf{(0.066)}&\textbf{(0.065)}\\
      &\multirow{2}{*}{LBFL}&1.716&96.289&0.979&\textbf{0.920}&0.950&0.967&0.581&0.642\\
      &&(0.870)&(5.362)&(0.048)&\textbf{(0.090)}&(0.051)&(0.101)&(0.145)&(0.126)\\
      &\multirow{2}{*}{LBFH}&\textbf{1.062}&\textbf{92.329}&0.995&0.905&\textbf{0.950}&\textbf{0.970}&0.871&0.887\\
      &&\textbf{(0.480)}&\textbf{(3.813)}&(0.022)&(0.134)&\textbf{(0.067)}&\textbf{(0.096)}&(0.089)&(0.078)\\
      \hline
      \multirow{6}{*}{$\bm{\beta}_{2}^{*}$}&\multirow{2}{*}{LFL}&6.180&91.536&\textbf{1.000}&0.482&0.871&\textbf{0.980}&0.847&0.868\\
      &&(1.205)&(3.603)&\textbf{(0.000)}&(0.447)&(0.112)&\textbf{(0.080)}&(0.097)&(0.083)\\
      &\multirow{2}{*}{LBFL}&1.742&88.395&0.977&0.930&0.966&0.977&0.598&0.657\\
      &&(0.791)&(3.967)&(0.037)&(0.104)&(0.039)&(0.085)&(0.125)&(0.107)\\
      &\multirow{2}{*}{LBFH}&\textbf{1.520}&\textbf{84.530}&0.997&\textbf{0.966}&\textbf{0.989}&0.970&\textbf{0.881}&\textbf{0.895}\\
      &&\textbf{(0.914)}&\textbf{(3.086)}&(0.015)&\textbf{(0.110)}&\textbf{(0.029)}&(0.107)&\textbf{(0.091)}&\textbf{(0.079)}\\
      \hline
   	\end{tabular}
      \end{table}

   \begin{table}[H]
   	\centering
      \caption{MSE (standard deviation), EL, PV, PZV, AV, PF, PNF, and AF for Case 2. Bold font indicates the smallest value of MSE and EL and the largest value of PV, PZV, AV, PF, PNF, and AF before rounding among LFL, LBFL, and LBFH.}
      \label{tab:case2}
      \begin{tabular}{c|ccccccccc}
         \hline
         \multicolumn{2}{c}{}&MSE&EL&PV&PZV&AV&PF&PNF&AF\\
         \multicolumn{2}{c}{}&(sd)&(sd)&(sd)&(sd)&(sd)&(sd)&(sd)&(sd)\\
         \hline
      \multirow{6}{*}{$\bm{\beta}_{1}^{*}$}&\multirow{2}{*}{LFL}&0.718&118.113&\textbf{1.000}&0.481&0.741&\textbf{1.000}&\textbf{0.898}&\textbf{0.914}\\
      &&(0.297)&(2.373)&\textbf{(0.000)}&(0.306)&(0.153)&\textbf{(0.000)}&\textbf{(0.083)}&\textbf{(0.070)}\\
      &\multirow{2}{*}{LBFL}&0.949&121.116&0.999&0.937&0.968&0.993&0.581&0.646\\
      &&(0.435)&(3.396)&(0.010)&(0.087)&(0.045)&(0.047)&(0.148)&(0.125)\\
      &\multirow{2}{*}{LBFH}&\textbf{0.381}&\textbf{116.649}&\textbf{1.000}&\textbf{0.985}&\textbf{0.993}&\textbf{1.000}&0.888&0.905\\
      &&\textbf{(0.194)}&\textbf{(2.220)}&\textbf{(0.000)}&\textbf{(0.046)}&\textbf{(0.023)}&\textbf{(0.000)}&(0.094)&(0.079)\\
      \hline 
      \multirow{6}{*}{$\bm{\beta}_{2}^{*}$}&\multirow{2}{*}{LFL}&5.350&80.432&\textbf{1.000}&0.364&0.841&\textbf{1.000}&0.846&0.871\\
      &&(0.790)&(2.902)&\textbf{(0.000)}&(0.424)&(0.106)&\textbf{(0.000)}&(0.078)&(0.066)\\
      &\multirow{2}{*}{LBFL}&1.860&79.103&0.982&0.942&0.972&0.980&0.561&0.627\\
      &&(0.894)&(4.442)&(0.031)&(0.118)&(0.040)&(0.080)&(0.138)&(0.117)\\
      &\multirow{2}{*}{LBFH}&\textbf{1.375}&\textbf{74.886}&0.994&\textbf{0.990}&\textbf{0.993}&0.987&\textbf{0.871}&\textbf{0.889}\\
      &&\textbf{(0.730)}&\textbf{(3.396)}&(0.019)&\textbf{(0.044)}&\textbf{(0.019)}&(0.066)&\textbf{(0.094)}&\textbf{(0.082)}\\
      \hline
   	\end{tabular}
   \end{table}
   
   \begin{table}[H]
   	\centering
      \caption{MSE (standard deviation), EL, PV, PZV, AV, PF, PNF, and AF for Case 3. Bold font indicates the smallest value of MSE and EL and the largest value of PV, PZV, AV, PF, PNF, and AF before rounding among LFL, LBFL, and LBFH.}
      \label{tab:case3}
      \begin{tabular}{c|ccccccccc}
         \hline
         \multicolumn{2}{c}{}&MSE&EL&PV&PZV&AV&PF&PNF&AF\\
         \multicolumn{2}{c}{}&(sd)&(sd)&(sd)&(sd)&(sd)&(sd)&(sd)&(sd)\\
         \hline
      \multirow{6}{*}{$\bm{\beta}_{1}^{*}$}&\multirow{2}{*}{LFL}&0.690&134.037&\textbf{1.000}&0.465&0.733&\textbf{1.000}&0.846&0.870\\
      &&(0.243)&(2.457)&\textbf{(0.000)}&(0.254)&(0.127)&\textbf{(0.000)}&(0.089)&(0.075)\\
      &\multirow{2}{*}{LBFL}&0.815&137.327&0.999&0.925&0.962&\textbf{1.000}&0.564&0.633\\
      &&(0.458)&(4.360)&(0.010)&(0.093)&(0.046)&\textbf{(0.000)}&(0.147)&(0.123)\\
      &\multirow{2}{*}{LBFH}&\textbf{0.297}&\textbf{132.313}&\textbf{1.000}&\textbf{0.982}&\textbf{0.991}&\textbf{1.000}&\textbf{0.883}&\textbf{0.902}\\
      &&\textbf{(0.167)}&\textbf{(2.760)}&\textbf{(0.000)}&\textbf{(0.046)}&\textbf{(0.023)}&\textbf{(0.000)}&\textbf{(0.086)}&\textbf{(0.072)}\\
      \hline
      \multirow{6}{*}{$\bm{\beta}_{2}^{*}$}&\multirow{2}{*}{LFL}&4.681&89.716&\textbf{1.000}&0.440&0.860&\textbf{1.000}&0.807&0.837\\
      &&(0.973)&(3.554)&\textbf{(0.000)}&(0.405)&(0.101)&\textbf{(0.000)}&(0.094)&(0.080)\\
      &\multirow{2}{*}{LBFL}&1.531&88.550&0.996&0.940&0.982&0.993&0.626&0.684\\
      &&(0.923)&(4.763)&(0.019)&(0.104)&(0.029)&(0.047)&(0.149)&(0.127)\\
      &\multirow{2}{*}{LBFH}&\textbf{1.030}&\textbf{84.256}&0.999&\textbf{0.980}&\textbf{0.994}&\textbf{1.000}&\textbf{0.904}&\textbf{0.919}\\
      &&\textbf{(0.657)}&\textbf{(3.792)}&(0.009)&\textbf{(0.060)}&\textbf{(0.016)}&\textbf{(0.000)}&\textbf{(0.077)}&\textbf{(0.065)}\\
      \hline
   	\end{tabular}
   \end{table}
   \begin{table}[H]
   	\centering
      \caption{MSE (standard deviation), EL, PV, PZV, AV, PF, PNF, and AF for Case 4. Bold font indicates the smallest value of MSE and EL and the largest value of PV, PZV, AV, PF, PNF, and AF before rounding among LFL, LBFL, and LBFH.}
      \label{tab:case4}
      \begin{tabular}{c|ccccccccc}
         \hline
         &MSE&EL&PV&PZV&AV&PF&PNF&AF\\
         &(sd)&(sd)&(sd)&(sd)&(sd)&(sd)&(sd)&(sd)\\
         \hline
      \multirow{2}{*}{LFL}&66.735&124.512&\textbf{0.998}&0.947&0.954&\textbf{0.890}&0.989&0.988\\
      &(0.984)&(3.965)&\textbf{(0.008)}&(0.061)&(0.051)&\textbf{(0.144)}&(0.005)&(0.005)\\
      \multirow{2}{*}{LBFL}&\textbf{42.765}&115.630&0.578&0.999&0.936&0.840&0.924&0.923\\
      &\textbf{(1.639)}&(11.928)&(0.068)&(0.001)&(0.010)&(0.157)&(0.013)&(0.013)\\
      \multirow{2}{*}{LBFH}&60.976&\textbf{109.058}&0.857&\textbf{1.000}&\textbf{0.978}&0.563&\textbf{0.999}&\textbf{0.994}\\
      &(1.855)&\textbf{(10.577)}&(0.151)&\textbf{(0.0004)}&\textbf{(0.023)}&(0.231)&\textbf{(0.002)}&\textbf{(0.004)}\\
      \hline
   	\end{tabular}
   \end{table}
   The results are summarized in Tables \ref{tab:case1-0}, \ref{tab:case1-0.5}, \ref{tab:case2}, \ref{tab:case3}, and \ref{tab:case4}. 
   LBFH gives the smallest ELs in all cases. 
   LBFH also gives the smallest MSEs in almost all cases except for Case 4. 
   These results show that LBFH outperformed LFL and LBFL in terms of the estimation and prediction accuracy. 
   In addition, LBFH achieves the largest AVs and AFs in most cases. 
   This shows that LBFH provides superior performance of variable selection and variable fusion compared to the other methods.
   Comparing LBFL to LFL, LBFL gives the smaller MSEs and ELs in more than half of the cases.
   LFL often gives larger PVs than LBFL, but LBFL gives larger PZVs and AVs in almost all cases.
   These results show that LBFL performs better than LFL in terms of the accuracy of estimation, prediction, and variable selection. 
   For the performance of variable fusion, PFs, PNFs, and AFs of LFL are competitive or larger than those of LBFL. 

\subsection{Application}
 \label{sec:application}
We applied our proposed methods LBFL and LBFH to the Wafer dataset, which was formatted in \citet{olszewski2001generalized} and analyzed in \citet{deng2014smt}. 
The dataset can be obtained from the UCR Time Series Classification Archive (\url{https://www.cs.ucr.edu/%7Eeamonn/time_series_data_2018}). 
Utilizing the dataset, the normal and abnormal etching processes of a wafer in semiconductor microelectronics manufacturing were classified based on time series data from six sensors (which monitor radio frequency forward power, radio frequency reflected power, chamber pressure, 405 nanometer emission, 520 nanometer emission, and direct current bias, respectively). 
Each time series data contains the value from one of the six sensors for one wafer and has length $p=152$. 
We labeled abnormal data as one and normal data as zero. 
The dataset contains $n=1{,}000$ training data and 6,164 test data. 
The abnormal data constitute 10.7\% of the training data and 12.1\% of the test data, meaning that the dataset has a large class imbalance.

We compared LBFL and LBFH to LFL. 
As in \citet{deng2014smt}, we selected the hyper-parameters $\lambda_1$ and $\lambda_2$ for LFL from four candidates, 0.05, 0.1, 0.3, and 0.5, by BIC. 
For LBFL and LBFH, the Gibbs sampler was run with 10,000 iterations, and then the first 6,000 samples were discarded as burn-in.

We evaluated the performance of LBFL, LBFH, and LFL by Area Under the ROC Curve (AUC). 
AUCs tend to be large in situations where the true positive rate is large when the false positive rate is small. The values of AUC are summarized in Table \ref{tab:wafer}. 
From Table \ref{tab:wafer}, LBFL gives the largest AUC.
Meanwhile, AUC is not enough to evaluate the performance of the model when there is a large class imbalance in the dataset.
In the case of such a large class imbalance, Area Under the Precision-Recall Curve (PR-AUC) is a more suitable indicator.
PR-AUCs tend to be large in situations where the true positive rate is large when the precision is large.
For the details of PR-AUC, we refer the reader to \citet{davis2006relationship}. 
The values of PR-AUC are also summarized in Table \ref{tab:wafer}, 
which shows that LBFH gives the largest PR-AUC.

\begin{figure}[H]
\centering
\begin{subfigmatrix}{1}
    \subfigure[LFL]{\includegraphics[height=6.03cm,width=17cm]{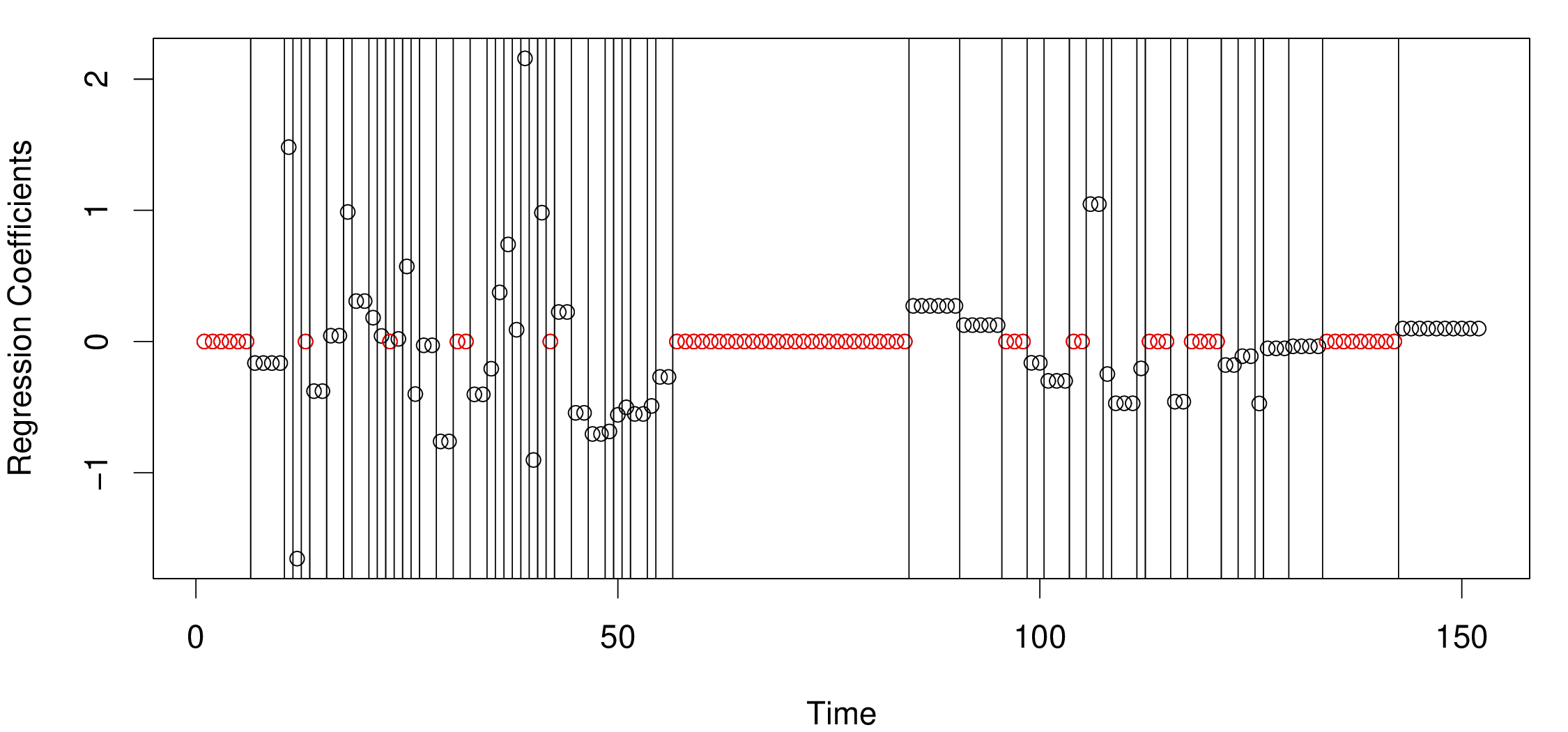}}
    \subfigure[LBFL]{\includegraphics[height=6.03cm,width=17cm]{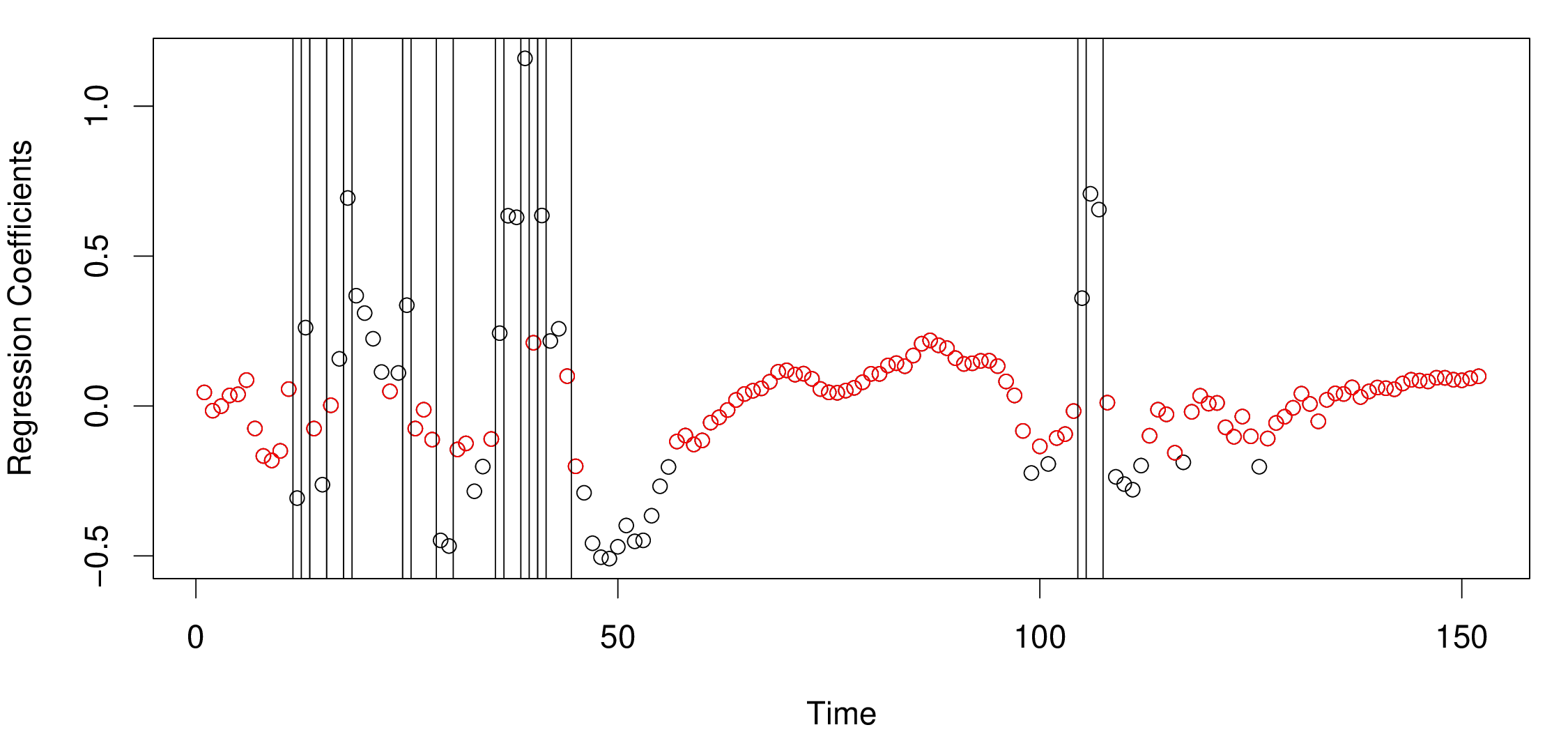}}
    \subfigure[LBFH]{\includegraphics[height=6.03cm,width=17cm]{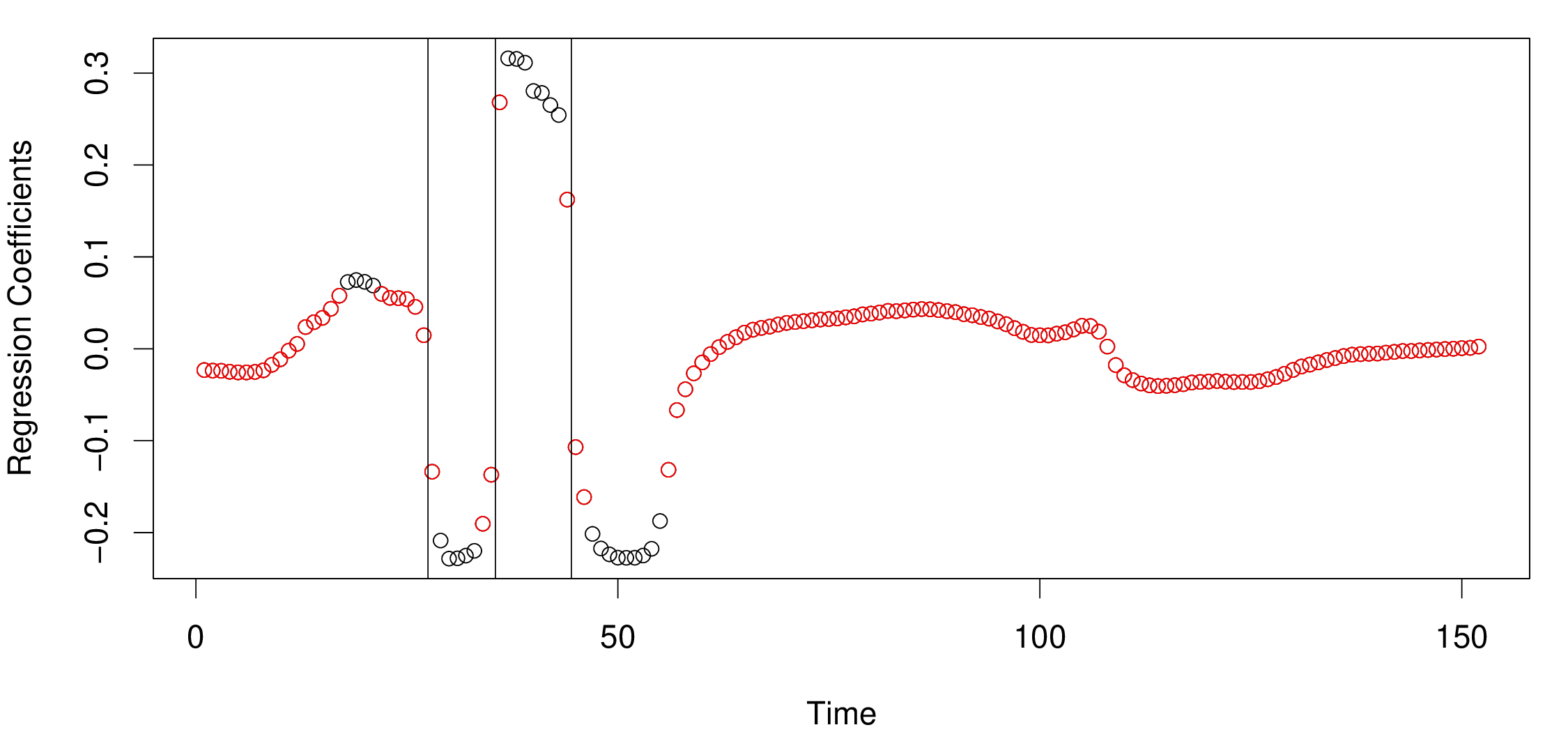}}
\end{subfigmatrix}
\caption{Estimated regression coefficients for LFL, LBFL, and LBFH. 
Dots indicate the values of the estimated regression coefficients.
For LFL, a red dot indicates a zero element of the estimated regression coefficients and a black vertical line indicates where a non-zero difference between adjacent estimated regression coefficients exists. 
For LBFL and LBFH, a red dot indicates an estimated regression coefficient whose MCMC samples give a 95\% credible interval including zero and a black vertical line indicates where a difference between adjacent regression coefficients whose MCMC samples give a 50\% credible interval not including zero exists. }
\label{fig:app_coef}
\end{figure}

As with the Monte Carlo simulations in Sec. \ref{sec:Monte Carlo}, we judged that the regression coefficient is estimated as zero if the corresponding 95\% credible interval includes zero for LBFL and LBFH. 
We also judged that the difference between adjacent regression coefficients is regarded as non-zero when the 50\% credible interval for the difference does not include zero.
The estimated regression coefficients are shown in Figure \ref{fig:app_coef}. 
From Figure \ref{fig:app_coef}, we observe that LFL estimated 60 regression coefficients as zero, whereas LBFL estimated the most as zero, at 133, and LBFH estimated the second most, at 127. 
Thus, the number of points which are considered to be unnecessary for the model is larger for LBFL and LBFH than for LFL, meaning that the former two make clearer which points are necessary for the prediction.


From Figure \ref{fig:app_coef}(c), we see that LBFH split the regression coefficients into four groups. 
The first group contains the 1st to 27th coefficients, the second contains the 28th to 35th coefficients, the third one the 36th to 44th, and the fourth one the 45th to 152nd.
On the other hand, from Figure \ref{fig:app_coef}(b), we see that LBFL split the regression coefficients into 21 groups, whereas from Figure \ref{fig:app_coef}(a), we see that LFL split the coefficients into 57 groups.
Thus, LBFH gave the smallest number of groups of variables, whereas LBFL gave the second smallest.
Based on the number of detected groups, LBFH seems to provide smoother and more simplified estimation than the other methods.
Many of the groups detected by LFL contain only one variable, 
meaning that LFL has weak performance regarding grouping multiple variables.

\begin{table}[H]
   	\centering
      \caption{AUC and PR-AUC for the Wafer dataset. 
      Bold font indicates the largest value among LFL, LBFL, and LBFH.}
      \label{tab:wafer}
      \begin{tabular}{c|cc}
         \hline
         &AUC&PR-AUC\\
         \hline
      LFL&0.880&0.615\\
      LBFL&\textbf{0.886}&0.592\\
      LBFH&0.864&\textbf{0.626}\\
      \hline
   	\end{tabular}
   \end{table}

\section{Conclusions}
\label{sec:conclusion}
We proposed a logistic regression model which enables variable selection by a Laplace prior and variable fusion by a Laplace prior or a horseshoe prior. 
Using a data-augmentation method with a P\'{o}lya-Gamma distribution and the hierarchical structure of the priors, we developed the Gibbs sampler. 
Through numerical studies, we showed that our proposed methods gave more accurate estimation and prediction than the existing method. 
Our proposed methods also outperformed the existing method in terms of variable selection and variable fusion without the selection of the hyper-parameters.

We used credible intervals for the evaluation of the performance of variable selection and variable fusion for Bayesian methods. 
It would be interesting to investigate the theoretical aspects of credible intervals for determining whether the target can be considered as zero or not. 
We leave that as future work.

\section*{Acknowledgements}
Y. K. was supported by JST, Establishment of University Fellowships towards the Creation of Science Technology Innovation, Grant Number JPMJFS2136. S. K. was supported by JSPS KAKENHI Grant Numbers JP23K11008, JP23H03352, and JP23H00809. Computational resources were provided by the Super Computer System, Human Genome Center, Institute of Medical Science, The University of Tokyo. The authors thank FORTE Science Communications (\url{https://www.forte-science.co.jp/}) for English language editing.

\bibliographystyle{spbasic}
\bibliography{papersanko}

\end{document}